\documentclass[12pt,a4paper]{article}

\usepackage[latin2]{inputenc}
\usepackage{epic}
\usepackage{latexsym}
\usepackage{amssymb,amsmath}

\def\beq{\begin{equation}}
\def\eeq{\end{equation}}
\def\beqa{\begin{eqnarray}}
\def\eeqa{\end{eqnarray}}

\def\nn{\nonumber}
\def\all{&&\!\!\!\!\!\!\!\!}
\def\d{{\rm d}}

\def\Tr{{\rm Tr}}

\def\si{\sigma}
\def\pa{\partial}

\def\Lie{{\pounds}_n}
\def\cT{{\cal T}}
\def\cL{{\cal L}}
\def\cTb{{\overline{\cal T}}}

\def\cR{{\cal R}}
\def\cRt{{\cal R_{**}^{**}}}
\def\Boxt{{\square_{*}^{*}}}
\def\Box{{\square}}

\begin{document}

\thispagestyle{empty}

%version: \today
\begin{flushright}
                     IFT--07--1
\end{flushright}
\vspace*{25mm}

\begin{center}
{\Large\bf Generalization of Einstein--Lovelock theory\\}
{\Large\bf to higher order dilaton gravity\\}
\vspace*{5mm}
\vspace*{5mm}
\end{center}
\vspace*{5mm} \noindent
\vskip 0.5cm
\centerline{\bf 
D. Konikowska, M. Olechowski}
\vskip 5mm
\centerline{\em Institute of Theoretical Physics, 
University of Warsaw}
\centerline{\em ul.\ Ho\.za 69, PL-00 681 Warsaw, Poland}
\vskip 3mm

\vskip 1cm

\centerline{\bf Abstract}
\vskip 3mm

A higher order theory of dilaton gravity is constructed
as a generalization of the Einstein--Lovelock 
theory of pure gravity. Its Lagrangian contains terms 
with higher powers of the Riemann tensor and of the first 
two derivatives of the dilaton. Nevertheless, the resulting 
equations of motion are quasi--linear in the second derivatives 
of the metric and of the dilaton. This property is crucial 
for the existence of brane solutions in the thin wall limit.
At each order in derivatives the contribution to the Lagrangian 
is unique up to an overall normalization.
Relations between symmetries of this theory and 
the $O(d,d)$ symmetry of the string--inspired models are discussed.

\vskip 3mm

%%%%%%%%%%%%%%%%%%%%%%%%%%%%%%%%%%%%%%%%%%%%%%%%%%%%%%%%%%%%%%%%%%%%%
%%%%%%%%%%%%%%%%%%%%%%%%%%%%%%%%%%%%%%%%%%%%%%%%%%%%%%%%%%%%%%%%%%%%%
\newpage
\section{Introduction}
%%%%%%%%%%%%%%%%%%%%%%%%%%%%%%%%%%%%%%%%%%%%%%%%%%%%%%%%%%%%%%%%%%%%%
%%%%%%%%%%%%%%%%%%%%%%%%%%%%%%%%%%%%%%%%%%%%%%%%%%%%%%%%%%%%%%%%%%%%%

The equations of motion in the Einstein theory of gravity
in 4 space--time dimensions are the most general divergence--free 
tensor (rank 2) equations bilinear 
in the first derivatives and linear in the second derivatives 
of the metric. They can be obtained from the Hilbert--Einstein 
action which is linear in the Riemann tensor. In more than 4 
space--time dimensions, this theory can be generalized to 
contain higher powers of the Riemann tensor in the action. 
The corresponding equations of motion involve higher powers 
of the first derivatives of the metric and are quasi--linear 
in the second derivatives (all terms are at most linear in 
the second derivatives, while multiplied by 
powers of the first derivatives). It has been shown that the contribution
to the action of a given order in the Riemann tensor is unique 
up to an overall normalization. The quadratic contribution 
is called the Gauss--Bonnet action or the Lanczos action 
\cite{La}. 
It has been generalized to higher orders by Lovelock \cite{Lo}. 
The quasi--linearity is a very important feature of the 
Einstein--Lovelock equations of motion.  
It guarantees that they can be formulated as a Cauchy problem 
with some constraints on the initial data \cite{Ch-Br}.
On the other hand, it is crucial for the existence of 
non--singular domain wall solutions in the thin wall 
limit. This problem for arbitrary order in derivatives was 
discussed in \cite{MeOl}. 
Many aspects of the Einstein--Lovelock gravity were discussed 
in the literature\footnote{
Quasi--linearity of the Einstein--Gauss--Bonnet theory 
was reviewed in \cite{DeMa}. A discussion of general 
quasi--linear differential equations can be found  
in \cite{CoHi}. For a review on brane--world gravity 
see eg.\ \cite{Ma}. For a discussion of the Lovelock 
gravity in the context of the equivalence of
the Palatini and metric formulations see eg.\ \cite{ExSh}.
}.

Higher derivative corrections to the gravity interactions 
are present in effective Lagrangians obtained from 
string theories. The first correction has exactly 
the form of the Gauss--Bonnet term \cite{Zw}, \cite{BoDe}. 
The lowest order dilaton interactions were added to the 
Gauss--Bonnet theory in \cite{BoDe2}.
However, the $\alpha'$ expansion in string theories predicts 
higher derivative corrections not only for the gravitational 
interactions, such corrections appear also 
for the dilaton. The effective action for the 
dilaton gravity with terms up to four derivatives 
was given in \cite{GrSl}, \cite{MeTs}.
The effective action with six derivatives was presented 
in \cite{MeTs2}, but its gravitational part has a form different 
from that of the corresponding Einstein--Lovelock action.

The dilaton gravity at the field theory level has been  
investigated by many authors. Some of them included 
also certain higher order corrections. Yet in most cases such
corrections were considered only for gravitational 
interactions and not for the dilaton. 
Some higher derivative corrections 
for both the dilaton and the gravitational interactions
were considered in 
\cite{MaRi}--\cite{TsSa} 
%\cite{MaRi}, \cite {NoOdOs}, \cite{Gi}, \cite{ChDaDu}, 
%\cite{Da}, \cite{AmChDa}, \cite{TsSa} 
(certain Riemann tensor combinations with dilaton dependent 
coefficients were analyzed in 
\cite{Ne}--\cite{NoOdSami}).
%\cite{Ne}, \cite{NoOdSa}, \cite{CaNe}, \cite{NoOdSami} ).
The terms predicted by superstrings up to four derivatives 
have also been considered in
\cite{BiChDaDu}--\cite{MaPa}.
%\cite{BiChDaDu}, \cite{JaMeOl}, \cite{MaPa}.

The purpose of the present work is to find a generalization 
of the lowest order dilaton gravity theory to an arbitrary 
order in derivatives. We start with the 
Einstein--Lovelock higher order gravity 
and couple it to the dilaton. There are many ways to do 
this but we are only interested in the theories where 
dilaton and gravity interactions are as similar to each
other as possible. Equations of motion in such a theory are 
presented in Section \ref{equations}. We begin with formulating 
the conditions which should be fulfilled by such equations. Most of 
them are simple generalizations of the conditions fulfilled by 
the Einstein--Lovelock equations of motion. One condition is
added in order to eliminate at least some of the possible theories
in which the dilaton interactions are not related to the gravitational 
ones. The equations of motion satisfying all those conditions
are constructed in Subsection \ref{construction}. 
It turns out that 
at each order those equations are unique up to a numerical 
normalization. Moreover, they can be obtained by the standard 
Euler--Lagrange procedure from the Lagrangian presented 
in Section \ref{lagrangian}. 
Section \ref{quasi} contains the proof that our
equations of motion are quasi--linear in the second
derivatives of both the metric and the dilaton. 
The relation between the gravity and the dilaton interactions 
is discussed in Section \ref{symmetries}.
We point out that the Lagrangian of our higher order dilaton 
gravity can be obtained in a simple way from the pure gravity 
Einstein--Lovelock Lagrangian. We also discuss the relation of the 
resulting theory to the $O(d,d)$ symmetric theories.
We conclude in Section \ref{conclusions}. The Appendix 
contains the explicit formulae for the Lagrangian and the 
equations of motion up to terms of the sixth order in 
derivatives.

%%%%%%%%%%%%%%%%%%%%%%%%%%%%%%%%%%%%%%%%%%%%%%%%%%%%%%%%%%%%%%%%%%%%%
%%%%%%%%%%%%%%%%%%%%%%%%%%%%%%%%%%%%%%%%%%%%%%%%%%%%%%%%%%%%%%%%%%%%%
\section{Equations of motion}
\label{equations}
%%%%%%%%%%%%%%%%%%%%%%%%%%%%%%%%%%%%%%%%%%%%%%%%%%%%%%%%%%%%%%%%%%%%%
%%%%%%%%%%%%%%%%%%%%%%%%%%%%%%%%%%%%%%%%%%%%%%%%%%%%%%%%%%%%%%%%%%%%%

%%%%%%%%%%%%%%%%%%%%%%%%%%%%%%%%%%%%%%%%%%%%%%%%%%%%%%%%%%%%%%%%%%%%%
\subsection{Notation}
\label{notation}
%%%%%%%%%%%%%%%%%%%%%%%%%%%%%%%%%%%%%%%%%%%%%%%%%%%%%%%%%%%%%%%%%%%%%

Let us start with introducing certain generalizations of the Kronecker 
delta and the trace operator which will be used later to make
the formulae more compact.
The generalized Kronecker delta is defined by
\beq
\delta_{i_1 i_2 \cdots i_n}^{j_1 j_2 \cdots j_n}
=
\det\left|
\begin{tabular}{cccc}
$\delta_{i_1}^{j_1}$ & $\delta_{i_2}^{j_1}$ & $\cdots$
& $\delta_{i_n}^{j_1}$\\
$\cdot$ & $\cdot$ & $\cdots$ & $\cdot$\\
$\cdot$ & $\cdot$ & $\cdots$ & $\cdot$\\
$\delta_{i_1}^{j_n}$ & $\delta_{i_2}^{j_n}$ & $\cdots$ 
& $\delta_{i_n}^{j_n}$\\
\end{tabular}
\right|\,,
\label{delta}
\eeq
and should be only employed when the spacetime dimensionality $D$ 
is sufficient: $D\ge n$.
Using this definition it is easy to prove some relations 
among Kronecker deltas of different order. For example:
\beq
\delta_{\mu\,i_1 i_2 \ldots i_n}^{\nu\,j_1 j_2 \ldots j_n}
=
\delta_{\mu}^{\nu}
\delta_{i_1 i_2 \ldots i_n}^{j_1 j_2 \ldots j_n}
-
\delta_{i_1}^{\nu}
\delta_{\mu\,\,i_2 \ldots i_n}^{j_1 j_2 \ldots j_n}
-
\delta_{i_2}^{\nu}
\delta_{i_1\mu\,\, \ldots i_n}^{j_1 j_2 \ldots j_n}
-\ldots
-
\delta_{i_n}^{\nu}
\delta_{i_1 i_2 \ldots\mu\,}^{j_1 j_2 \ldots j_n}
\,.
\label{delta_expand}
\eeq
The generalized Kronecker delta can be used to define the 
following trace--like linear mapping from tenors of rank $(n,n)$ 
into numbers
\beq
\cT(M)
=
\delta_{i_1 i_2 \cdots i_n}^{j_1 j_2 \cdots j_n}
M^{i_1 i_2 \cdots i_n}_{j_1 j_2 \cdots j_n}\,,
\label{cT}
\eeq
which reduces to the ordinary trace for $n=1$.
We will also employ an extension of this operation which maps
tensors of rank $(n,n)$ into tensors of rank $(1,1)$:
\beq
\cTb_{\!\mu}^{\,\nu}(M)
=
\delta_{\mu\, i_1 i_2 \cdots i_n}^{\nu\, j_1 j_2 \cdots j_n}
M^{i_1 i_2 \cdots i_n}_{j_1 j_2 \cdots j_n}\,.
\label{cTb}
\eeq
In the following we will often use $\cT$ and $\cTb$ 
evaluated for products of tensors. In order to clearly 
distinguish between tensors and their contracted 
counterparts, we will use $*$ indices to 
indicate the rank of a tensor. For example, 
$\cRt$ denotes the rank $(2,2)$ Riemann tensor, and 
$\Boxt\phi$ denotes the rank $(1,1)$ second derivative of 
the dilaton, while $\cR$ is the Ricci scalar and $\Box\phi$ 
the D'Alembertian acting on the dilaton.
Thus, for example,
\beq
\cT\left(\left(\cRt\right)^2\left(\Boxt\phi\right)^2\right)
=
\delta_{\rho_1\rho_2\rho_3\rho_4\rho_5\rho_6}
^{\si_1\si_2\si_3\si_4\si_5\si_6}
\,\cR_{\si_1\si_2}^{\rho_1\rho_2} \,\cR_{\si_3\si_4}^{\rho_3\rho_3}
\,\Box^{\rho_5}_{\si_5}\phi \,\Box^{\rho_6}_{\si_6}\phi\,,
\eeq
where we used the notation 
$\cR_{\si_1\si_2}^{\rho_1\rho_2}={\cR^{\rho_1\rho_2}}_{\si_1\si_2}$
and $\Box^{\rho}_{\si}\phi=\nabla^\rho\pa_\si\phi$
to make the formula more compact.
It is easy to see that the sequence of tensors appearing in the
product argument of $\cT$ is not important. Changing such an order 
is equivalent to interchanging the appropriate columns of indices in the
generalized Kronecker delta. On the other hand, interchanging 
two such columns of indices is equivalent to interchanging  
the corresponding 2 rows and 2 columns in the determinant in
Definition (\ref{delta}). Each interchange of two columns 
(or two rows) changes the sign of the determinant, hence an even number
of interchanges leaves the determinant unchanged.

%%%%%%%%%%%%%%%%%%%%%%%%%%%%%%%%%%%%%%%%%%%%%%%%%%%%%%%%%%%%%%%%%%%%%
\subsection{Conditions} 
\label{conditions}
%%%%%%%%%%%%%%%%%%%%%%%%%%%%%%%%%%%%%%%%%%%%%%%%%%%%%%%%%%%%%%%%%%%%%

Now we want to construct the $n$--th order dilaton gravity 
equations of motion. They are to be of the form
\beqa
T_{\mu\nu}^{(n)}=0\,,
\nn\\[2pt]
\qquad\qquad
W^{(n)}=0\,,
\label{eom}
\eeqa
where the tensor $T_{\mu\nu}^{(n)}$ and the scalar $W^{(n)}$ 
satisfy the following conditions
\renewcommand{\labelenumi}{(\roman{enumi})}
\begin{enumerate}
\item
They are combinations of terms with exactly $2n$ derivatives 
acting on the metric tensor $g_{\mu\nu}$ and on the dilaton 
field $\phi$. There are no derivatives higher than second 
acting on one object;
\item 
Tensor $T_{\mu\nu}^{(n)}$ is symmetric in its indices;
\item
The covariant derivative of the tensor 
is proportional to the scalar: \\
$\nabla_\nu T^{\nu(n)}_\mu={\rm const}\cdot(\pa_\mu\phi) W^{(n)}$ 
(the energy--momentum tensor is covariantly conserved if the 
dilaton equation of motion is fulfilled).
\end{enumerate}

It is clear that the above conditions are not sufficient to 
determine something which could be regarded as an extension
of the higher order gravity theory to the dilaton gravity case. 
For example, all the above conditions are fulfilled by the 
Einstein--Gauss--Bonnet gravity with only the 
lowest order terms for the dilaton.
We are interested in a theory where the dilaton and the 
metric are treated in a more symmetric way. It is not obvious 
how such a symmetry should be defined, because it ought to relate 
a scalar to a second rank tensor. Or, more precisely, it is supposed 
to relate the first and second derivatives of the scalar field
to the Riemann tensor and its contractions.
A simple observation concerning the gravity part is that 
it contains even--rank tensors only. On the other hand, the 
first derivative of a scalar is a rank--1 tensor. Hence one 
can expect that in a gravity--dilaton symmetric theory, 
the first derivative of the dilaton appears only as 
a 0--rank tensor: $g^{\mu\nu}\pa_\mu\phi\pa_\nu\phi$. 
However, the feature mentioned above is not invariant under change 
of variables. Thus, we should specify in which frame 
it is fulfilled. The theory which relates dilaton to 
gravity is the string theory so the string frame 
seems to be a natural choice. Hence our last condition reads:
\begin{enumerate}
\addtocounter{enumi}{3}
\item
In the string--like frame, in which the pure gravity term is 
multiplied by $\exp(-\phi)$, the first order derivatives of the 
dilaton appear in the combination 
$(\pa_\mu\phi)(\pa^\mu\phi)$ only.
\end{enumerate}
The relation of this condition to the $O(d,d)$ symmetry 
present in many string--inspired theories will be discussed 
in Section \ref{symmetries}.

%%%%%%%%%%%%%%%%%%%%%%%%%%%%%%%%%%%%%%%%%%%%%%%%%%%%%%%%%%%%%%%%%%%%%
\subsection{Construction} 
\label{construction}
%%%%%%%%%%%%%%%%%%%%%%%%%%%%%%%%%%%%%%%%%%%%%%%%%%%%%%%%%%%%%%%%%%%%%

We start our construction with a term in $T^{\nu(n)}_\mu$ where
all $2n$ derivatives act on the metric tensors. The only pure gravity 
tensor satisfying Conditions (i)--(iii)
(with $W^{(n)}=0$)
is, up to normalization, equal to the $n$-th order Lovelock 
tensor \cite{Lo}. Because of Condition (iv), it is 
most natural to work in the frame in which the gravity 
term is multiplied by $\exp(-\phi)$. Consequently, the tensor $T_\mu^{\nu(n)}$ 
starts with
\beq
T_\mu^{\,\nu\,(n)}
=
-2^{-(n+1)}e^{-\phi}
\delta_{\mu\,\rho_1\ldots\rho_{2n}}
^{\nu\,\si_1\ldots\si_{2n}}
R^{\rho_1\rho_2}_{\si_1\si_2}\cdots
R^{\rho_{2n-1}\rho_{2n}}_{\si_{2n-1}\si_{2n}}+\ldots
\label{T1}
\eeq
The reason for such a normalization will be explained 
in the next section. Calculating the divergence 
of (\ref{T1}), we get
\beq
\nabla_\nu T_\mu^{\,\nu\,(n)}
=
2^{-(n+1)}e^{-\phi}
(\pa_\nu\phi)
\delta_{\mu\,\rho_1\ldots\rho_{2n}}^{\nu\,\si_1\ldots\si_{2n}}
R^{\rho_1\rho_2}_{\si_1\si_2}\cdots
R^{\rho_{2n-1}\rho_{2n}}_{\si_{2n-1}\si_{2n}}+\ldots
\label{dT1}
\eeq
The above term is produced when the derivative acts on $e^{-\phi}$
(derivatives of the Riemann tensor do not contribute due to the 
Bianchi identity).

The first term in $T_\mu^{\nu(n)}$ shown explicitly in (\ref{T1}) can 
not be the only one. The reason is that the r.h.s.\ of 
(\ref{dT1}) is not a product of $\pa_\mu\phi$ and a scalar, 
as Condition (iii) requires. 
Using Eq.\ (\ref{delta_expand}) we can rewrite the r.h.s.\ of 
(\ref{dT1}) as a combination of $(2n+1)$ terms. The one containing 
the first term from the r.h.s.\ of (\ref{delta_expand}) 
is of the desired form but the remaining $2n$ terms have 
different structures of the index contractions.
It turns out that 
similar terms are also present in 
the following covariant derivative
\beqa
&&
\nabla_\nu \left[e^{-\phi}
\delta_{\mu\,\rho_1\ldots\rho_{2n-1}}
^{\nu\,\si_1\ldots\si_{2n-1}}
R^{\rho_1\rho_2}_{\si_1\si_2}\cdots
R^{\rho_{2n-3}\rho_{2n-2}}_{\si_{2n-3}\si_{2n-2}}
\Box^{\rho_{2n-1}}_{\si_{2n-1}}\phi\right]
=
\nn\\[4pt]
&&\qquad\qquad=
- e^{-\phi}(\pa_\nu\phi)
\delta_{\mu\,\rho_1\ldots\rho_{2n-1}}^{\nu\,\si_1\ldots\si_{2n-1}}
R^{\rho_1\rho_2}_{\si_1\si_2}\cdots
R^{\rho_{2n-3}\rho_{2n-2}}_{\si_{2n-3}\si_{2n-2}}
\Box^{\rho_{2n-1}}_{\si_{2n-1}}\phi
+\qquad
\nn\\[4pt]
&&
\qquad\qquad\,\,\,\,\,+
e^{-\phi}
\delta_{\mu\,\rho_1\ldots\rho_{2n-1}}
^{\nu\,\si_1\ldots\si_{2n-1}}
R^{\rho_1\rho_2}_{\si_1\si_2}\cdots
R^{\rho_{2n-3}\rho_{2n-2}}_{\si_{2n-3}\si_{2n-2}}
\left(\nabla_\nu\nabla^{\rho_{2n-1}}\pa_{\si_{2n-1}}\phi\right)\,.
\label{dRbox}
\eeqa
The second term on the r.h.s.\ may be rewritten as
\beqa
&
e^{-\phi}
\delta_{\mu\,\rho_1\ldots\rho_{2n-1}}
^{\nu\,\si_1\ldots\si_{2n-1}}
R^{\rho_1\rho_2}_{\si_1\si_2}\cdots
R^{\rho_{2n-3}\rho_{2n-2}}_{\si_{2n-3}\si_{2n-2}}
\left(-\frac12\delta_{\rho_{2n}}^{\si_{2n}}
R_{\si_{2n-1}\nu}^{\rho_{2n-1}\rho_{2n}}\pa_{\si_{2n}}\phi\right)
=
\nn\\[4pt]
&\qquad\qquad
=
-\frac12 e^{-\phi}
(\pa_\nu\phi)
\delta_{\rho_{2n}}^{\nu}
\delta_{\rho_1\ldots\rho_{2n-1}\mu}
^{\si_1\ldots\si_{2n-1}\si_{2n}}
R^{\rho_1\rho_2}_{\si_1\si_2}\cdots
R^{\rho_{2n-3}\rho_{2n-2}}_{\si_{2n-3}\si_{2n-2}}
R_{\si_{2n-1}\si_{2n}}^{\rho_{2n-1}\rho_{2n}}\,,
\qquad
\label{dRbox2}
\eeqa
where in the last step we interchanged the names
of the contracted indices $\nu$ and $\si_{2n}$ and rearranged 
the indices in the generalized Kronecker delta.
A term exactly of this structure must be added to (\ref{dT1}) 
in order to obtain an expression proportional to $\pa_\mu\phi$.
From Eqs.\ (\ref{delta_expand}) and (\ref{dT1}) 
it follows that the coefficient should be equal to $(-n2^{-n})$ instead
of the $(-1/2)$ present in (\ref{dRbox2}). This fixes the coefficient of 
the term in $T_\mu^{\nu(n)}$ which contains $(n-1)$ Riemann tensors 
and one second derivative of the dilaton. Now we know the first 
two terms of the tensor $T_\mu^{\nu(n)}$. Using the notation introduced in
(\ref{cT}) and (\ref{cTb}), they can be written as:
\beq
T_\mu^{\,\nu\,(n)}
=
-2^{-(n+1)}e^{-\phi}
\cTb_\mu^{\,\nu}\left(\left(\cRt\right)^n\right)
-2^{-(n-1)}ne^{-\phi}
\cTb_\mu^{\,\nu}\left(\left(\cRt\right)^{(n-1)}\Boxt\phi\right)
+\ldots
\label{T2}
\eeq
Their covariant derivative reads
\beqa
\nabla_\nu T_\mu^{\,\nu\,(n)}
=\all
2^{-(n+1)}e^{-\phi}(\pa_\mu\phi)
\cTb\left(\left(\cRt\right)^n\right)
\nn\\
\all+
2^{-(n-1)}ne^{-\phi}(\pa_\nu\phi)
\cTb_\mu^{\,\nu}\left(\left(\cRt\right)^{(n-1)}\Boxt\phi\right)
+\ldots
\label{dT2}
\eeqa
The first term has the structure required by Condition (iii)
and determines the first term of the scalar equation of 
motion\footnote
{
up to an overall normalization. The choice of the relative 
normalizations of $T_{\mu\nu}^{(n)}$ and $W^{(n)}$  
shall become clear when the Lagrangian is introduced 
in Section \ref{lagrangian}.
} $W^{(n)}$.
However, the second term in (\ref{dT2})
is not of the appropriate structure. It means that some additional terms, 
whose covariant derivatives are 
products of $(n-1)$ Riemann tensors with one second derivative 
of the dilaton, are necessary in 
$T_\mu^{\nu(n)}$. Two such terms are possible:
\beq
c_3e^{-\phi}\cTb_\mu^{\,\nu}
\left(\left(\cRt\right)^{(n-2)}\left(\Boxt\phi\right)^2\right)
+c_4
e^{-\phi}\cTb_\mu^{\,\nu}\left(\left(\cRt\right)^{(n-1)}\right)
(\pa\phi)^2\,.
\label{c3c4}
\eeq
However, it is not enough to have terms with appropriate powers of 
the Riemann tensor and the dilaton, because their covariant 
divergences must contain the correct combinations of the generalized 
Kronecker deltas. To check whether this is possible, we calculate 
the covariant divergence of (\ref{c3c4}).
When the derivative acts on $\Boxt\phi$ in the first 
term in (\ref{c3c4}), it gives an additional Riemann tensor 
multiplied by $\pa\phi$ and a pair of new indices. Those new
indices are contracted with just one ordinary Kronecker delta and are not 
under the overall antisymmetrization. Similarly, when the covariant 
derivative acts on $(\pa\phi)^2$ in the second term in 
(\ref{c3c4}), it gives the second derivative of the dilaton 
multiplied by $\pa\phi$ and a pair of new indices. 
Those two covariant derivatives should combine with the second
term on the r.h.s.\ of (\ref{dT2}) to give an expression 
proportional to $\pa_\mu\phi$. This fixes the
numerical coefficients $c_3$ and $c_4$. 
The explicit calculation gives $c_3=-2^{(2-n)}n(n-1)$, 
$c_4=2^{-n}n$. Thus, we have found the first four terms of $T_\mu^{\nu(n)}$:
\beqa
T_\mu^{\,\nu\,(n)}
=
-2^{-(n+1)}e^{-\phi}
\all\left[
\cTb_\mu^{\,\nu}\left(\left(\cRt\right)^n\right)
+4n
\cTb_\mu^{\,\nu}\left(\left(\cRt\right)^{(n-1)}\Boxt\phi\right)
\right.
+
\nn\\[4pt]
\all\,\,
+8n(n-1)\cTb_\mu^{\,\nu}
\left(\left(\cRt\right)^{(n-2)}\left(\Boxt\phi\right)^2\right)
+
\nn\\[4pt]
\all\,\,
\left.-2n
\cTb_\mu^{\,\nu}\left(\left(\cRt\right)^{(n-1)}\right)
(\pa\phi)^2
\right]+\ldots
\label{T4}
\eeqa
The covariant divergence of those terms reads
\beqa
\nabla_\nu T_\mu^{\,\nu\,(n)}
=\all
\pa_\mu\phi\,
\left\{
2^{-(n+1)}e^{-\phi}\left[
\cT\left(\left(\cRt\right)^n\right)
+4n
\cT\left(\left(\cRt\right)^{(n-1)}\Boxt\phi\right)
\right]\right\}
\nn\\[4pt]
\all+
\pa_\nu\phi\,2^{(2-n)}n(n-1)e^{-\phi}\cTb_\mu^{\,\nu}
\left(\left(\cRt\right)^{(n-2)}\left(\Boxt\phi\right)^2\right)
\nn\\[4pt]
\all-
\pa_\nu\phi\,2^{-n}n
e^{-\phi}\cTb_\mu^{\,\nu}\left(\left(\cRt\right)^{(n-1)}\right)
(\pa\phi)^2
+\ldots
\label{dT4}
\eeqa
The terms in the curly bracket above are the first two
terms of the scalar $W^{(n)}$ we are looking for.

Equation (\ref{dT4}) shows that the procedure of finding
$T_\mu^{\nu(n)}$ and $W^{(n)}$ must be continued. The last two terms
on the r.h.s.\ of (\ref{dT4}) do not have the required form,
so more terms must be added to $T_\mu^{\nu(n)}$. From the steps 
described so far, it should be clear that each of such new 
terms must contain exactly 3 (first or second
order) derivatives of the dilaton. There are two such terms:
\beq
c_5e^{-\phi}\cTb_\mu^{\,\nu}\left(\left(\cRt\right)^{(n-3)}
\left(\Boxt\phi\right)^3\right) 
+c_6
e^{-\phi}\cTb_\mu^{\,\nu}\left(\left(\cRt\right)^{(n-2)}
\Boxt\phi\right)(\pa\phi)^2\,.
\eeq
The coefficients $c_5$ and $c_6$ can be fixed in the same 
way as $c_3$ and $c_4$.

This procedure can be continued step by step for the terms 
containing higher and higher powers of the dilaton field with 
the derivatives acting on it. Eventually, one obtains the term with 
the maximal number of dilaton fields, namely 
$c\,e^{-\phi}\delta_\mu^\nu\left[(\pa\phi)^2\right]^n$.
This is the first term in $T_\mu^{\nu(n)}$, the covariant
derivative of which need not to be corrected by contributions from
any additional terms. This covariant derivative reads
\beq
\nabla_\nu\left[e^{-\phi}\delta_\mu^\nu\left[(\pa\phi)^2\right]^n
\right]
=
-(\pa_\mu\phi) e^{-\phi}\left[(\pa\phi)^2\right]^n
+2ne^{-\phi}(\nabla_\mu\pa^\si\phi)(\pa_\si\phi)
\left[(\pa\phi)^2\right]^{(n-1)}\,.
\label{dphi2n}
\eeq
The second term on the r.h.s.\ is used to cancel some unwanted 
part of \\
$\nabla_\mu\left[e^{-\phi}\cTb_\mu^{\,\nu}(\Boxt\phi)
\left[(\pa\phi)^2\right]^{(n-1)}\right]$, which fixes 
$c$ to be equal to $\frac12(-1)^{(n+1)}$. 
The first term on the r.h.s.\ of 
(\ref{dphi2n}) has already the required structure of the product
of $\pa_\mu\phi$ and a scalar. Thus, the procedure can stop
here.

The above iterative procedure gives $T_{\mu\nu}^{(n)}$ and $W^{(n)}$ 
satisfying all the four imposed conditions.
The resulting gravitational and dilaton equations of motion
can be written in the following relatively simple form:%
\beqa
T^{(n)}_{\mu\nu}
\all
=
-\frac12 e^{-\phi}\sum_{a=0}^n\sum_{b=0}^{n-a}
\frac{2^{b-a}n!}{a!b!(n-a-b)!}
\,\cTb_{\mu\nu}\left(\left(\cRt\right)^a
\left(\Boxt\phi\right)^b\right)
\left(-(\partial\phi)^2\right)^{n-a-b}=
\nn\\[4pt]
\all
=0\,,
\label{Tn}
\eeqa
\beqa
W^{(n)}
\all
=
-e^{-\phi}\sum_{a=0}^n\sum_{b=0}^{n-a}
\frac{2^{b-a}n!}{a!b!(n-a-b)!}
\,\cT\!\left(\left(\cRt\right)^a
\left(\Boxt\phi\right)^b\right)
\left(-(\partial\phi)^2\right)^{n-a-b}=
\nn\\[4pt]
\all=0\,.
\label{Wn}
\eeqa

The existence of $T_{\mu\nu}^{(n)}$ and $W^{(n)}$  
is a non--trivial result, because in our iterative procedure
there are more conditions than 
available constants. A priori it could happen that there 
were no solutions other than a trivial one with vanishing
$T_{\mu\nu}^{(n)}$ and $W^{(n)}$. However, the solution exists and is unique
up to an overall normalization. Hence any 
dilaton gravity equations of motion, satisfying Conditions 
(i)--(iv), which contain at least
one term present in (\ref{Tn}) and (\ref{Wn}) must also contain
all the other terms with uniquely determined coefficients.

%%%%%%%%%%%%%%%%%%%%%%%%%%%%%%%%%%%%%%%%%%%%%%%%%%%%%%%%%%%%%%%%%%%%%
%%%%%%%%%%%%%%%%%%%%%%%%%%%%%%%%%%%%%%%%%%%%%%%%%%%%%%%%%%%%%%%%%%%%%
\section{Lagrangian}
\label{lagrangian}
%%%%%%%%%%%%%%%%%%%%%%%%%%%%%%%%%%%%%%%%%%%%%%%%%%%%%%%%%%%%%%%%%%%%%
%%%%%%%%%%%%%%%%%%%%%%%%%%%%%%%%%%%%%%%%%%%%%%%%%%%%%%%%%%%%%%%%%%%%%

It is interesting to check whether the equations of motion
constructed in Section 2 
can be obtained from some $D$--dimensional 
action. In such case, $T^{(n)}_{\mu\nu}$ and $W^{(n)}$ 
would satisfy
\beq
\delta_{g^{\mu\nu}}S^{(n)}
=
\delta_{g^{\mu\nu}} \int\d^D x \sqrt{-g}\,\cL^{(n)}
=
\int \d^D x \sqrt{-g}\, T^{(n)}_{\mu\nu}\delta g^{\mu\nu}
,
\label{var_g}
\eeq
\beq
\delta_{\phi}S^{(n)}
=
\delta_{\phi} \int\d^D x \sqrt{-g}\,\cL^{(n)}
=
\int \d^D x \sqrt{-g}\, W^{(n)}\delta\phi
\,.
\label{var_phi}
\eeq
It turns out that indeed the equations of motion (\ref{Tn}) 
and (\ref{Wn}) can be obtained from the action with the 
Lagrangian density given by
\beq
\cL^{(n)}
=
e^{-\phi}\sum_{a=0}^n\sum_{b=0}^{n-a}
\frac{2^{b-a}n!}{a!b!(n-a-b)!}
\,\cT\!\left(\left(\cRt\right)^a
\left(\Boxt\phi\right)^b\right)
\left(-(\partial\phi)^2\right)^{n-a-b}
\,.
\label{Ln}
\eeq

It is important to underline that for Conditions (i)--(iv) 
not to be violated, the terms coming 
from the $n$-th Lagrangian can appear only in the 
space--times with dimensionality $ D \geq 2n $.
Moreover, one should be careful when calculating 
(\ref{var_g}) for $D=2n$, as the generalized Kronecker delta 
(\ref{delta}) can not be employed in (\ref{Tn})
for the term of the highest order in the Riemann tensor.
The coefficient of that term should be replaced with
\beq
\delta_{\mu\,\rho_1\rho_2\ldots\rho_{2n}}^{\nu\,\si_1\si_2\ldots\si_{2n}}
\,\,{\underset{D=2n}{\longrightarrow}}\,\,
\delta_{\mu}^{\nu}
\delta_{\rho_1\rho_2\ldots\rho_{2n}}^{\si_1\si_2\ldots\si_{2n}}
-
\delta_{\rho_1}^{\nu}
\delta_{\mu\,\,\rho_2\ldots\rho_{2n}}^{\si_1\si_2\ldots\si_{2n}}
-
\delta_{\rho_2}^{\nu}
\delta_{\rho_1\mu\,\,\ldots\rho_{2n}}^{\si_1\si_2\ldots\si_{2n}}
-\ldots
-
\delta_{\rho_{2n}}^{\nu}
\delta_{\rho_1\rho_2\ldots\mu\,}^{\si_1\si_2\ldots\si_{2n}}
\,.
\eeq

Now we can comment on the overall 
normalization of the tensors $T^{(n)}_{\mu\nu}$. 
The reason for this particular normalization 
is that the term $e^{-\phi}\cR^n$ 
(with $\cR$ being the Ricci scalar) appears in the 
Lagrangian with the coefficient 1. 
This corresponds to the standard normalizations of 
the Hilbert--Einstein and Gauss--Bonnet Lagrangians.

Proving that the equations of motion derived from the Lagrangian
(\ref{Ln}) really have the form (\ref{eom}) with $T^{(n)}_{\mu\nu}$ 
and $W^{(n)}$ as given in (\ref{Tn}) and (\ref{Wn})
is a straightforward but quite tedious calculation. 
One of the reasons is that apparently several integrations by parts
are required. This can be somewhat simplified 
if one observes that not all those integrations by parts
have to be performed explicitly. In case of (\ref{var_phi}), 
the reason is as follows. Under the integral (\ref{var_phi}) 
there are first (second) derivatives of $\delta\phi$ coming 
from the variation of the first (second) derivatives of the dilaton.
In general, the terms containing second derivatives of $\delta\phi$
should be integrated by parts twice. However,
one can notice that the result of a single integration 
and the terms containing the first derivatives of $\delta\phi$ 
cancel each other exactly.

The situation is a little bit more complicated 
in case of the gravitational equation of motion.
Under the integral (\ref{var_g}), 
there are second derivatives of $\delta g^{\mu\nu}$ coming
from the variation of the Riemann tensor and 
first derivatives of  $\delta g^{\mu\nu}$ coming from the 
variation of the second covariant derivative of the dilaton.
Similarly as in the case of the dilaton equation of motion,
the terms containing second derivatives of $\delta g^{\mu\nu}$
have to be integrated by parts only once. And although 
the cancellation of the resulting terms is not
complete this time, only some residual integration by parts
has to be performed additionally.

Of course, the Lagrangian density (\ref{Ln}) is not unique.
First, one can rewrite $\cL^{(n)}$ changing the variables 
$g_{\mu\nu}$ and $\phi$. Second, 
one can add to $\cL^{(n)}$ any total divergence without changing
the resulting equations of motion. However, the form given
in Eq.\ (\ref{Ln}) is especially simple and interesting.
It is very similar to the form of $T^{(n)}_{\mu\nu}$ and $W^{(n)}$. 
The energy momentum tensor $T^{(n)}_{\mu\nu}$ can be obtained from 
$\cL^{(n)}$ by replacing the generalized trace $\cT$ with
its tensor extension $\cTb_{\mu\nu}$ and multiplying the result
by $-1/2$. In case of the dilaton equation of motion, the 
analogous relation is even simpler: $W^{(n)}=-\cL^{(n)}$.

We were not able to find any other similarly simple form
of the Lagrangian by adding total derivative terms or by
changing the variables. For example, we examined the form 
of the Lagrangian and of the equations of motion in the
Einstein--like frame in which the common factor $e^{-\phi}$ 
is absorbed by a suitable Weyl transformation. The results are 
very complicated and will not be presented here. One of the
reasons for such complications is that the Weyl transformation 
depends on the dimensionality $D$ of the space--time. 
Thus, many different functions of $D$ appear in the 
Einstein frame, while there is no explicit dependence 
on $D$ in our string--like frame.

%%%%%%%%%%%%%%%%%%%%%%%%%%%%%%%%%%%%%%%%%%%%%%%%%%%%%%%%%%%%%%%%%%%%%
%%%%%%%%%%%%%%%%%%%%%%%%%%%%%%%%%%%%%%%%%%%%%%%%%%%%%%%%%%%%%%%%%%%%%
\section{Quasi-linearity}
\label{quasi}
%%%%%%%%%%%%%%%%%%%%%%%%%%%%%%%%%%%%%%%%%%%%%%%%%%%%%%%%%%%%%%%%%%%%%
%%%%%%%%%%%%%%%%%%%%%%%%%%%%%%%%%%%%%%%%%%%%%%%%%%%%%%%%%%%%%%%%%%%%%

It is easy to show that the equations of motion (\ref{Tn})--(\ref{Wn})
are quasi--linear in the second derivatives of the metric
and the dilaton. 
Let us introduce in the $D$--dimensional space--time
a $(D-1)$--dimensional hypersurface $\Sigma$ defined by its unit 
normal vector $n^\mu$. The metric induced at this hypersurface 
is given by
\beq
h_{\mu\nu}=g_{\mu\nu}-\frac{n_\mu n_\nu}{n^2}
\,, 
\label{metric}
\eeq
where $n^{2}=n_\rho n^\rho$. 
The components of the $D$--dimensional Riemann tensor $\cRt$ 
corresponding to the full metric $g_{\mu\nu}$ can be expressed as
\beqa
\cR_{\mu\nu}^{\rho\si}
=
R_{\mu\nu}^{\rho\si}
\all
-n^{-2}\left(
2 K_{[\mu}^{[\rho} K_{\nu]}^{\si]}
+4 n^{[\rho}D_{[\mu}K_{\nu]}^{\si]}
+ 4 n_{[\mu}D^{[\rho}K_{\nu]}^{\si]}
\right)
\nn\\[4pt]
\all
+n^{-4}\left(
4 n_{[\mu}n^{[\rho}K^{\si]}_{|\tau|} K^{\tau}_{\nu]}
-4 n_{[\mu}n^{[\rho}\Lie K_{\nu]}^{\si]}
\right)
\,,
\label{R}
\eeqa
where: $R$ is the $(D-1)$--dimensional Riemann tensor 
corresponding to the induced metric $h_{\mu\nu}$;
$K$ is the extrinsic curvature given by
\beq
K_{\mu\nu}=\frac12 \Lie h_{\mu\nu}
\,;
\eeq
$D_\mu$ is the covariant derivative with respect to the induced metric 
$h_{\mu\nu}$; $\Lie$ is the Lie derivative along the 
vector field $n^\mu$. 
Similarly we can write the $D$--dimensional second 
covariant (with respect to the metric $g_{\mu\nu}$) derivative of 
the dilaton
\beqa
\nabla_\mu \nabla_\nu \phi
=
D_\mu D_\nu \phi 
\all 
+ n^{-2}\left(
K_{\mu\nu} \Lie \phi
+ 2 n_{(\mu} D_{\nu)} \Lie \phi 
- 2 n_{(\mu} K_{\nu)}^\tau D_\tau \phi
\right)
\nn\\[4pt]
\all
+ n^{-4}\, n_\mu n_\nu 
\left(\Lie^2 \phi 
- \left(n^\rho\nabla_\rho n^\tau\right)\nabla_\tau \phi\right)
.
\label{Box}
\eeqa

We want to check how the second Lie derivatives of the metric 
$h_{\mu\nu}$ (present in $\Lie K_{\mu\nu}$) and of the dilaton
$\phi$ appear in the equations of motion (\ref{Tn}) and (\ref{Wn}). 
Such second derivatives are present in (\ref{R}) 
and (\ref{Box}) but in both cases they are multiplied 
by coefficients bilinear in the vector $n$. 
After substituting the decompositions (\ref{R}) and (\ref{Box}) 
into (\ref{Tn}) and (\ref{Wn}), one can immediately see 
that, due to the antisymmetrization present in $T_{\mu\nu}^{(n)}$
and $W^{(n)}$, the equations of motion contain terms at most
bilinear in $n$. Thus, the equations of motion 
(\ref{Tn}) and (\ref{Wn}) contain terms at most linear
in the second Lie derivatives of $h_{\mu\nu}$ and $\phi$.

We have shown that the equations of motion are quasi--linear 
in the second Lie derivatives ``perpendicular'' to the 
hypersurface $\Sigma$.
This quasi--linearity has very important consequences. 
For $\Sigma$ with a time--like $n$, this allows us to define 
a standard Cauchy problem with the
initial conditions (values and first Lie derivatives of 
$h_{\mu\nu}$ and $\phi$) given at $\Sigma$. 
For a space--like $n$, the quasi--linearity is necessary 
to have non--singular brane solutions even in the thin 
wall limit.

%%%%%%%%%%%%%%%%%%%%%%%%%%%%%%%%%%%%%%%%%%%%%%%%%%%%%%%%%%%%%%%%%%%%%
%%%%%%%%%%%%%%%%%%%%%%%%%%%%%%%%%%%%%%%%%%%%%%%%%%%%%%%%%%%%%%%%%%%%%
\section{Symmetries}
\label{symmetries}
%%%%%%%%%%%%%%%%%%%%%%%%%%%%%%%%%%%%%%%%%%%%%%%%%%%%%%%%%%%%%%%%%%%%%
%%%%%%%%%%%%%%%%%%%%%%%%%%%%%%%%%%%%%%%%%%%%%%%%%%%%%%%%%%%%%%%%%%%%%

The equations of motion presented in Section \ref{equations} 
were obtained assuming some kind of symmetry between 
the metric and the dilaton. Now we are in a position to 
investigate such a symmetry in more detail. 

It is quite amazing 
that the Lagrangian (\ref{Ln}) as well as the equations of motion
(\ref{Tn}) and (\ref{Wn}) can be expressed as functions of 
$n$--th perfect ``power'' of one simple $n$--independent quantity.
Namely:
\beqa
\cL^{(n)}
\all=
-W^{(n)}=
e^{-\phi}\cT\left[
\left(\frac12\cRt 
\,\oplus\,
 2\Boxt\phi 
\,\oplus\, 
(-1)\left(\pa\phi\right)^2\right)^{\!n\,}
\right],
\label{Lnc}
\\[2pt]
T^{(n)}_{\mu\nu}
\all=
-\frac12e^{-\phi}\cTb_{\mu\nu}\left[
\left(\frac12\cRt 
\,\oplus\, 
2\Boxt\phi 
\,\oplus\, 
(-1)\left(\pa\phi\right)^2\right)^{\!n\,}
\right].
\label{Tnc}
\eeqa
One can treat these equations as just a new notation allowing us
to rewrite the double sums from (\ref{Tn}), (\ref{Wn}) 
and (\ref{Ln}) in a compact 
way\footnote{
One could say that Eqs.\ (\ref{Lnc}) and (\ref{Tnc}) make no 
sense because they contain a sum of tensors of different ranks. 
To make this mathematically sensible, we should consider a simple
sum of spaces of tensors of a given rank. Then the tensors in 
(\ref{Lnc}) and (\ref{Tnc}) should be understood as elements of 
such a sum space with all but one components set to zero. Finally, 
the generalized traces $\cT$ and $\cTb_{\mu\nu}$ should be
further extended in such a way that when acting on an element
of this big space they give the result being the sum of 
generalized traces of all components.
}.
Yet, on the other hand, it helps to show that the action and the 
equations of motion depend on some combinations of the 
dilaton derivatives and tensors obtained from the metric only. 
In each round parenthesis in Eqs.\ (\ref{Lnc}) and (\ref{Tnc}), 
there are the rank--4 Riemann tensor, the rank--2 tensor of 
the second derivatives of the dilaton and the rank--0 tensor 
built from the first derivatives of the dilaton:
\beq
\frac12\cRt 
\,\oplus\, 
2\Boxt\phi 
\,\oplus\, 
(-1)\left(\pa\phi\right)^2
.
\label{tensors}
\eeq
All those tensors are under the generalized traces 
$\cT$ and $\cTb_{\mu\nu}$. 
Some of the terms present in these mappings 
contain traces of the tensors from (\ref{tensors}).
There are two different rank--2 tensors coming from (\ref{tensors}).
The first is just $\Boxt\phi$. The second is the Ricci tensor 
$\cR_*^*$ which can be obtained from the Riemann tensor by 
contraction of its two indices. There are four different 
ways to contract one pair of indices in $\cRt$, thus in the 
final result the rank--2 tensors appear always in the 
combination $(2\cR_*^* + 2\Boxt\phi)$.
There are three different scalars originating from (\ref{tensors}):
$\left(\pa\phi\right)^2$, $\Box\phi$ and the curvature scalar $\cR$.
There are two different constructions giving $\cR$, so the final 
results depend on a single following scalar combination: 
$\cR+2\Box\phi-\left(\pa\phi\right)^2$.

The above observation allows us to relate our dilaton gravity 
equations to the corresponding equations in the pure 
Einstein--Lovelock gravity:
\beqa
\cL^{(n)}
\all=
-W^{(n)}=
e^{-\phi}\cL^{(n)}_{\rm E-L}
\Big[\cRt\,,\left(\cR_*^*+\Boxt\phi\right)
,\left(\cR+2\Box\phi-\left(\pa\phi\right)^2\right)
\Big]
,
\label{Ln_EL}
\\[4pt]
T^{(n)}_{\mu\nu}
\all=
e^{-\phi}\left(T^{(n)}_{\rm E-L}\right)_{\mu\nu}
\Big[g_{\mu\nu}\,,\cRt\,,\left(\cR_*^*+\Boxt\phi\right)
,\left(\cR+2\Box\phi-\left(\pa\phi\right)^2\right)
\Big]
.
\label{Tn_EL}
\eeqa
The recipe for the higher order dilaton gravity can be as follows:
Start with the higher order pure gravity Einstein--Lovelock theory.
Write the Lagrangian density $\cL^{(n)}_{\rm E-L}$ 
and the equations of motion $(T^{(n)}_{\rm E-L})_{\mu\nu}$ 
in terms of the Riemann tensor, Ricci tensor and the curvature 
scalar by performing all internal (within a given Riemann tensor)  
contractions of indices. Then make the substitutions:
\beqa
\cR_\rho^\si 
\all\rightarrow 
\left[\cR_\rho^\si + \Box_\rho^\si \phi\right]\,,
\\[4pt]
\cR 
\all\rightarrow 
\left[\cR + 2\Box\phi -\left(\pa\phi\right)^2 \right]\,.
\eeqa
Finally, multiply the result by $\exp(-\phi)$. The dilaton equation 
of motion, absent in the pure gravity case, is simply $\cL^{(n)}=0$.

It occurs that the form of the Lagrangian and the tensor 
$T_{\mu\nu}^{(n)}$ given in (\ref{Ln_EL}) and (\ref{Tn_EL}) 
is very closely related to the string $O(d,d)$ 
symmetry\footnote{
For a review on $O(d,d)$ symmetry, see e.g.\ \cite{MaSc} 
and the references therein.
}. To show this, we consider the $D$--dimensional 
block--diagonal metric of the form
\beq
g_{\mu\nu}
=
\left(
\begin{tabular}{cc}
$\tilde g_{\alpha\beta}$ & 0\\
0 & $G_{mn}$
\end{tabular}
\right)\,,
\eeq
where $\alpha,\beta=1,\ldots,(D-d)$; $m,n=(D-d+1),\ldots,D$.  
We assume that the metric components $\tilde g_{\alpha\beta}$ and 
$G_{mn}$ and the dilaton field $\phi$ do not depend on the 
last $d$ coordinates $x^m$. In such a case, we obtain the following 
expressions for the second derivatives of the dilaton
\beqa
\Box_\alpha^\beta\phi
=\all
\tilde\Box_\alpha^\beta\phi
\,,
\label{OddBoxab}
\\[2pt]
\Box_m^n\phi
=\all
\frac12\left(G^{-1}\pa_\alpha G\right)_m^n
\pa^\alpha\phi
\,,
\label{OddBoxmn}
\\[2pt]
\Box\phi
=\all
\tilde\Box\phi
+\frac12\left(\pa_\alpha\ln\det G\right) 
\pa^\alpha\phi
\,,
\label{OddBox}
\eeqa
and for the Ricci tensor and the curvature scalar
\beqa
\cR_\alpha^\beta
=\all
\tilde \cR_\alpha^\beta 
- \frac12\tilde\Box_\alpha^\beta\ln\det G
- \frac14\Tr\left[G^{-1}\left(\pa_\alpha G\right) 
G^{-1}\left(\pa^\beta G\right)\right]
\,,
\label{OddRab}
\\[2pt]
\cR_m^n
=\all
-\frac14\left(\pa_\alpha\ln\det G\right)
\left(G^{-1}\pa^\alpha G\right)_m^n
-\frac12\left(G^{-1}\tilde\Box G\right)_m^n
\nn\\[2pt]
\all
+\frac12\left[G^{-1}\left(\pa_\alpha G\right) 
G^{-1}\left(\pa^\alpha G\right)\right]_m^n
\,,
\label{OddRmn}
\\[2pt]
\cR
=\all
\tilde \cR
-\frac14\left(\pa_\alpha\ln\det G\right)
\left(\pa^\alpha\ln\det G\right)
-\tilde\Box\ln\det G
\nn\\[2pt]
\all
-\frac14\Tr\left[G^{-1}\left(\pa_\alpha G\right) 
G^{-1}\left(\pa^\alpha G\right)\right]
\,,
\label{OddR}
\eeqa
where tilde denotes quantities related to the $(D-d)$--dimensional
metric $\tilde g_{\alpha\beta}$, 
$G$ should be understood as a $d \times d$ matrix (and not its determinant)
and $\Tr$ and $\det$ are the trace and the determinant 
(acting on $d \times d$ matrices).

A necessary condition for the $O(d,d)$ symmetry is that the dilaton
field $\phi$ appear only in the $O(d,d)$ invariant combination
\begin{equation}
\Phi=\phi-\frac12\ln\det G\,.
\label{Phi}
\end{equation}
Hence any derivative of the dilaton $\phi$ must be accompanied 
by an appropriate derivative of $[\ln\det G]$.
It is easy to see that there are only three  
combinations of Eqs.\ (\ref{OddBoxab})--(\ref{OddR}) 
and the first derivatives of $\phi$  
which depend on $\phi$ and  $[\ln\det G]$ only through the
combination $\Phi$: 
\beqa
\cR + 2\Box\phi -\left(\pa\phi\right)^2 
=\all
\tilde\cR
+ 2\tilde\Box\Phi
- \pa_\alpha\Phi\pa^\alpha\Phi
%\nn\\[2pt]
%&&
-\frac14 \Tr\left[G^{-1}\left(\pa_\alpha G\right) 
G^{-1}\left(\pa^\alpha G\right)\right],
\label{Odd1}
\\[2pt]
\cR_\alpha^\beta + \Box_\alpha^\beta \phi
=\all
\tilde\cR_\alpha^\beta
+ \tilde\Box_\alpha^\beta\Phi
-\frac14 \Tr\left[G^{-1}\left(\pa_\alpha G\right) 
G^{-1}\left(\pa^\beta G\right)\right],
\label{Odd2}
\\[2pt]
\cR_m^n + \Box_m^n \phi
=\all
\frac12\left(\pa_\alpha\Phi\right)\left(G^{-1}\pa^\alpha G\right)_m^n
-\frac12\tilde\nabla_\alpha\left(G^{-1}\pa^\alpha G\right)_m^n\,.
\label{Odd3}
\eeqa
These are exactly the combinations which, together with 
the Riemann tensor with uncontracted indices, appear 
in the formulation given in Eqs.\ (\ref{Ln_EL}) and (\ref{Tn_EL}).
Hence the higher derivative contributions to the dilaton
gravity theory found in the present paper fulfill the 
necessary condition for the $O(d,d)$ symmetry formulated 
before Eq.\ (\ref{Phi}).
This does not mean yet that our theory is a part of some 
$O(d,d)$ symmetric theory. One should check whether 
all terms depending on $G$ other than $[\ln\det G]$
form only $O(d,d)$ invariant combinations. Actually, one can 
calculate that it is really the case for $n=1$ and $n=2$.
The lowest order theory was analyzed from this point of 
view for the first time in \cite{Odd1}. 
Our second order Lagrangian $\cL^{(2)}$ differs from the one 
found in \cite{Me} (for a vanishing tensor field $H$) 
by some total derivatives only. Thus, for $n=1,2$ the equations
of motion presented in Section \ref{equations} are the 
same as the dilaton and gravity part of the equations 
obtained as appropriate approximations from the superstring 
theories. The relation to the $O(d,d)$ symmetry for 
$n>2$ will be discussed elsewhere \cite{MeOlOdd}.

The above discussion shows that Condition (iv) from Section 
\ref{conditions} can be treated as
a necessary one for the dilaton gravity model to be part
of some $O(d,d)$ symmetric theory. The reason is that there 
are no $O(d,d)$ invariant expressions containing the first 
derivatives of the dilaton other than the combination  
$(\pa_\mu\phi)(\pa^\mu\phi)$.

%%%%%%%%%%%%%%%%%%%%%%%%%%%%%%%%%%%%%%%%%%%%%%%%%%%%%%%%%%%%%%%%%%%%%
%%%%%%%%%%%%%%%%%%%%%%%%%%%%%%%%%%%%%%%%%%%%%%%%%%%%%%%%%%%%%%%%%%%%%
\section{Conclusions}
\label{conclusions}
%%%%%%%%%%%%%%%%%%%%%%%%%%%%%%%%%%%%%%%%%%%%%%%%%%%%%%%%%%%%%%%%%%%%%
%%%%%%%%%%%%%%%%%%%%%%%%%%%%%%%%%%%%%%%%%%%%%%%%%%%%%%%%%%%%%%%%%%%%%

We have generalized the Einstein--Lovelock theory by adding  
interactions with the dilaton. The corresponding Einstein 
and dilaton 
equations of motion can be written as series in the number 
of derivatives acting on the fields:
\beqa
T_{\mu\nu}
=\all
\sum_{n} \kappa_n T_{\mu\nu}^{(n)}=0
\,,
\label{EoMT}
\\[2pt]
W
=\all
\sum_{n} \kappa_n W^{(n)}=0
\,.
\label{EoMW}
\eeqa
The $n$--th contributions $T_{\mu\nu}^{(n)}$ and $W^{(n)}$ 
are sums of terms 
containing products of the Riemann tensor and the first and second 
derivatives of the dilaton field. There are $2n$ derivatives 
in each such term. We have found the most general  
equations of motion satisfying Conditions (i)--(iv) 
given in Section \ref{conditions}. The first three conditions are 
the standard properties of the dilaton gravity theories. 
The last one was added in order to find the theories in 
which the dilaton and the metric are treated, as much as possible, 
on the same footing. Accordingly, we assumed 
that the rank--1 tensor containing the first derivatives of 
the dilaton can appear only in the scalar combination 
$(\pa_\mu\phi)(\pa^\mu\phi)$, as there is no way to build an odd--rank 
tensor from the metric and the Riemann tensor. It is necessary 
to specify the frame 
in which such a condition is to be fulfilled. We have chosen the
string frame where the $n$--th order term from the 
Einstein--Lovelock theory is multiplied by $e^{-\phi}$. 
The reason is quite simple: symmetries relating the dilaton and
the metric are present in string--motivated theories.

We have shown that at each order $T_{\mu\nu}^{(n)}$ and $W^{(n)}$ 
are unique up to a normalization. General expressions for 
$T_{\mu\nu}^{(n)}$ and $W^{(n)}$ for arbitrary $n$ are given
in Section \ref{construction}. The explicite formulae for 
$n\le3$ are presented in the Appendix. 
It occurs that the higher order dilaton gravity equations of motion 
have properties similar to those of the pure 
Einstein--Lovelock gravity. Namely:
\begin{itemize}
\item
There is an upper limit on the number of terms in 
(\ref{EoMT})--(\ref{EoMW}) 
which can be non--zero. For a $D$--dimensional space--time it is 
given by the inequality $2n\le D$ (the corresponding limit 
for pure gravity is $2n < D$)
\item
The equations of motion are quasi--linear in the second derivatives.
This allows us to treat them as a standard Cauchy initial conditions 
problem. It is crucial also for the existence of brane--type 
solutions in the thin wall limit.
\end{itemize}
There is also another very interesting feature of those equations. 
The form of the scalar and Einstein equations is very similar 
when written with the help of the generalized Kronecker delta. 
The tensor $T_{\mu\nu}^{(n)}$ can be obtained from the 
scalar $W^{(n)}$ simply by adding a pair of extra indices 
$\mu$ and $\nu$ to each generalized Kronecker delta and 
dividing by 2.

Our dilaton gravity equations of motion can be obtained from 
an appropriate Lagrangian. Of course, such a Lagrangian can be 
determined only up to some total derivatives. However, we have 
found that there is one particularly interesting form of it:
\beq
\cL=-W
\,.
\eeq
Moreover, this Lagrangian is related in a simple way to the 
Einstein--Lovelock one (the same is true also for the 
gravitational equations of motion). First, one has to 
write the Einstein--Lovelock Lagrangian as a function 
of the Riemann tensor, the Ricci tensor and the curvature scalar 
by performing all internal (within the same Riemann tensor) 
contractions of indices. Next, one should replace the curvature 
scalar with the combination $\cR+2\Box\phi-(\pa\phi)^2$, 
and the Ricci tensor with $\cR_\rho^\si+\Box_\rho^\si$. The result
is the dilaton gravity Lagrangian.

The property that the Lagrangian can be written in terms of 
only three tensors: one scalar $\cR+2\Box\phi-(\pa\phi)^2$,
one rank--2 tensor $\cR_\rho^\si+\Box_\rho^\si$ and the
rank--4 Riemann tensor is quite important. We have shown that 
this is a necessary condition for the dilaton gravity to
be a part of any string motivated theory with the $O(d,d)$ 
symmetry. It turns out that for $n=1,2$ it is also a sufficient one. 
The contributions $\cL^{(1)}$ and $\cL^{(2)}$ to our Lagrangian 
are, up to total derivatives, the same as those found from 
demanding the $O(d,d)$ symmetry \cite{Odd1}, \cite{Me}. 
It would be interesting to investigate the relation of 
$\cL^{(n)}$ to string theories for $n>2$ \cite{MeOlOdd}.

Most of the interesting features of the Lagrangian and the 
equations of motion are visible in the string frame only. 
The theory looks more complicated in other frames. For example, 
in the most often used Einstein frame there are no simple 
relations between tensors built from the metric and from 
the dilaton derivatives and also many coefficients become 
explicitly $D$--dependent. The advantages of the string frame 
should not be surprising. For example, much more explicit 
solutions in the lowest order dilaton gravity were found 
in the string frame \cite{MeOldg} than in the Einstein one 
(discussions concerning the relation between the string and 
the Einstein frames are reviewed in \cite{FaGuNa}).
Our results show that the string frame is the most 
convenient one to investigate dilaton gravity also at 
higher orders.

%%%%%%%%%%%%%%%%%%%%%%%%%%%%%%%%%%%%%%%%%%%%%%%%%%%%%%%%%%%%%%%%%%%%%
%%%%%%%%%%%%%%%%%%%%%%%%%%%%%%%%%%%%%%%%%%%%%%%%%%%%%%%%%%%%%%%%%%%%%

%%%%%%%%%%%%%%%%%%%%%%%%%%%%%%%%%%%%%%%%%%%%%%%%%%%%%%%%%%%%%%%%%%%%%
%%%%%%%%%%%%%%%%%%%%%%%%%%%%%%%%%%%%%%%%%%%%%%%%%%%%%%%%%%%%%%%%%%%%%
\section*{Acknowledgments}

The work of D.K.\ was partially supported by the EC Project 
MTKD-CT-2005-029466 ``Particle Physics and Cosmology: the Interface'' 
and by the Polish MEiN grant 1 P03B 099 29 for years 2005-2007.
M.O.\ acknowledges partial support from the EU 6th Framework
Program MRTN-CT-2004-503369 ``Quest for Unification''
and from the Polish MNiSW grant N202 176 31/3844 for years 2006-2008.
M.O.\ thanks for hospitality experienced at Institute of Theoretical 
Physics of Heidelberg University where part of this work has been done.

%%%%%%%%%%%%%%%%%%%%%%%%%%%%%%%%%%%%%%%%%%%%%%%%%%%%%%%%%%%%%%%%%%%%%
%%%%%%%%%%%%%%%%%%%%%%%%%%%%%%%%%%%%%%%%%%%%%%%%%%%%%%%%%%%%%%%%%%%%%
\section*{Appendix}
\renewcommand{\theequation}{A.\arabic{equation}}
\setcounter{equation}{0}
%%%%%%%%%%%%%%%%%%%%%%%%%%%%%%%%%%%%%%%%%%%%%%%%%%%%%%%%%%%%%%%%%%%%%
%%%%%%%%%%%%%%%%%%%%%%%%%%%%%%%%%%%%%%%%%%%%%%%%%%%%%%%%%%%%%%%%%%%%%

The dilaton gravity Lagrangian and the corresponding equations 
of motion can be written as a series in the number of 
derivatives
\beqa
\cL
=\all
\sum_{n=0}^{[D/2]} \kappa_n \cL^{(n)}
\,,
\\[2pt]
T_\mu^\nu
=\all
\sum_{n=0}^{[D/2]} \kappa_n T_\mu^{\nu(n)}=0
\,,
\eeqa
and the dilaton equation of motion $W = -\cL = 0$. \\
The 0--th order terms correspond to the cosmological constant:
\beqa
e^{\phi}\cL^{(0)}
=1 \all
\,,
\\[2pt]
e^{\phi}T_\mu^{\nu(0)}
=\all
-\frac12\delta_\mu^\nu
\,.
\eeqa
The 1--st order contribution is the standard Einstein gravity 
interacting with the dilaton:
\beqa
e^{\phi}\cL^{(1)}
=\all
\cR + 2\Box\phi -\left(\pa\phi\right)^2 ,
\\[2pt]
e^{\phi}T_\mu^{\nu(1)}
=\all
-\frac12 \delta_\mu^\nu e^{\phi}\cL^{(1)} 
+ \Big( \cR_\mu^\nu + \Box_\mu^\nu\phi \Big) . 
\eeqa
The next two orders are given by the following expressions:
\beqa
e^{\phi}\cL^{(2)}
=\all
\Big(e^{\phi}\cL^{(1)}\Big)^2
-4 \Big(\cR_{\rho_1}^{\rho_2} + \Box_{\rho_1}^{\rho_2} \phi\Big)
   \Big(\cR_{\rho_2}^{\rho_1} + \Box_{\rho_2}^{\rho_1} \phi\Big)
+ \cR^{\rho_2\rho_4}_{\rho_1\rho_3}\cR^{\rho_1\rho_3}_{\rho_2\rho_4}\,,
\\[2pt]
e^{\phi}T_\mu^{\nu(2)}
=\all 
-\frac12 \delta_\mu^\nu e^{\phi}\cL^{(2)} 
+2 \Big( \cR_\mu^\nu + \Box_\mu^\nu\phi \Big) e^{\phi}\cL^{(1)}
-4 \Big( \cR_\mu^\rho + \Box_\mu^\rho\phi \Big)
   \Big( \cR^\nu_\rho + \Box^\nu_\rho\phi \Big)
\qquad
\nn\\[2pt]
\all
-4 \cR_{\mu\rho_1}^{\nu\rho_2} 
   \Big( \cR^{\rho_1}_{\rho_2} + \Box^{\rho_1}_{\rho_2}\phi \Big)
+2 \cR^{\rho_1\rho_3}_{\mu\;\rho_2}\cR^{\nu\;\rho_2}_{\rho_1\rho_3}\,,
%\eeqa
%
%
%\beqa
\\[2pt]
e^{\phi}\cL^{(3)}
=\all 
3\Big(e^{\phi}\cL^{(2)}\Big)\Big(e^{\phi}\cL^{(1)}\Big)
-2\Big(e^{\phi}\cL^{(1)}\Big)^3
\nn\\[2pt]
\all
+16 \Big(\cR_{\rho_1}^{\rho_2} + \Box_{\rho_1}^{\rho_2} \phi\Big)
    \Big(\cR_{\rho_2}^{\rho_3} + \Box_{\rho_2}^{\rho_3} \phi\Big)
    \Big(\cR_{\rho_3}^{\rho_1} + \Box_{\rho_3}^{\rho_1} \phi\Big)
\nn\\[2pt]
\all
+24 \Big(\cR_{\rho_1}^{\rho_2} + \Box_{\rho_1}^{\rho_2} \phi\Big)
    \Big(\cR_{\rho_3}^{\rho_4} + \Box_{\rho_3}^{\rho_4} \phi\Big)
     \cR_{\rho_2\rho_4}^{\rho_1\rho_3}
-24 \Big(\cR_{\rho_1}^{\rho_2} + \Box_{\rho_1}^{\rho_2} \phi\Big)
    \cR^{\rho_1\rho_5}_{\rho_3\rho_4}
    \cR_{\rho_2\rho_5}^{\rho_3\rho_4}
\nn\\[2pt]
\all
-8 \cR_{\rho_1\rho_3}^{\rho_2\rho_4}
   \cR_{\rho_2\rho_5}^{\rho_1\rho_6}
   \cR_{\rho_4\rho_6}^{\rho_3\rho_5}      
+2 \cR_{\rho_1\rho_3}^{\rho_2\rho_4}
   \cR_{\rho_5\rho_6}^{\rho_1\rho_3}
   \cR_{\rho_2\rho_4}^{\rho_5\rho_6}\,,  
\eeqa
\beqa
%\\[2pt]
e^{\phi}T_\mu^{\nu(3)}
=\all
-\frac12 \delta_\mu^\nu e^{\phi}\cL^{(3)} 
+3 \Big( \cR_\mu^\nu + \Box_\mu^\nu\phi \Big) e^{\phi}\cL^{(2)}
-12\cR\Big(\cR_{\mu}^{\rho}+\Box_{\mu}^{\rho}\phi\Big)
\Big(\cR_{\rho}^{\nu}+\Box_{\rho}^{\nu}\phi\Big)
\nn\\[2pt]
\all
-12\cR\cR_{\mu\,\rho_1}^{\nu\rho_2}
\Big(\cR_{\rho_2}^{\rho_1}+\Box_{\rho_2}^{\rho_1}\phi\Big)
+6\cR\cR_{\mu\,\,\rho_2}^{\rho_1\rho_3}\cR_{\rho_1\rho_3}^{\nu\,\rho_2}
\nn\\[2pt]
\all
+24\Big(\cR_{\mu}^{\rho_1}+\Box_{\mu}^{\rho_1}\phi\Big)
\Big(\cR_{\rho_2}^{\nu}+\Box_{\rho_2}^{\nu}\phi\Big)
\Big(\cR_{\rho_1}^{\rho_2}+\Box_{\rho_1}^{\rho_2}\phi\Big)
\nn\\[2pt]
\all
+24\cR_{\mu\,\,\rho_1}^{\nu\rho_2}
\Big(\cR_{\rho_1}^{\rho_3}+\Box_{\rho_1}^{\rho_3}\phi\Big)
\Big(\cR_{\rho_3}^{\rho_2}+\Box_{\rho_3}^{\rho_2}\phi\Big)
+24\cR_{\mu\,\,\rho_1}^{\nu\rho_2}\cR_{\rho_2\rho_3}^{\rho_1\rho_4}
\Big(\cR_{\rho_4}^{\rho_3}+\Box_{\rho_4}^{\rho_3}\phi\Big)
\nn\\[2pt]
\all
+24\cR_{\mu\,\,\rho_2}^{\rho_1\rho_3}
\Big(\cR_{\rho_1}^{\nu}+\Box_{\rho_1}^{\nu}\phi\Big)
\Big(\cR_{\rho_3}^{\rho_2}+\Box_{\rho_3}^{\rho_2}\phi\Big)
-12\cR_{\mu\,\,\rho_2}^{\rho_1\rho_3}\cR_{\rho_1\rho_3}^{\rho_4\rho_2}
\Big(\cR_{\rho_4}^{\nu}+\Box_{\rho_4}^{\nu}\phi\Big)
\nn\\[2pt]
\all
+24\cR_{\rho_1\rho_2}^{\nu\,\rho_3}
\Big(\cR_{\mu}^{\rho_1}+\Box_{\mu}^{\rho_1}\phi\Big)
\Big(\cR_{\rho_3}^{\rho_2}+\Box_{\rho_3}^{\rho_2}\phi\Big)
-12\cR_{\rho_1\rho_2}^{\nu\,\rho_3}\cR_{\rho_4\rho_3}^{\rho_1\rho_2}
\Big(\cR_{\mu}^{\rho_4}+\Box_{\mu}^{\rho_4}\phi\Big)
\nn\\[2pt]
\all
-24\cR_{\mu\,\,\rho_2}^{\rho_1\rho_3}\cR_{\rho_1\rho_4}^{\nu\,\rho_2}
\Big(\cR_{\rho_3}^{\rho_4}+\Box_{\rho_3}^{\rho_4}\phi\Big)
-12\cR_{\mu\,\,\rho_2}^{\rho_1\rho_3}\cR_{\rho_4\rho_2}^{\nu\,\rho_3}
\Big(\cR_{\rho_3}^{\rho_4}+\Box_{\rho_3}^{\rho_4}\phi\Big)
\nn\\[2pt]
\all
-12\cR_{\mu\,\,\rho_1}^{\nu\rho_2}
\cR_{\rho_3\rho_4}^{\rho_1\rho_5}
\cR_{\rho_2\rho_5}^{\rho_3\rho_4}
+6\cR_{\mu\,\,\rho_2}^{\rho_1\rho_3}\cR_{\rho_4\rho_5}^{\nu\,\rho_2}
\cR_{\rho_1\rho_3}^{\rho_4\rho_5}
-24\cR_{\mu\,\,\rho_2}^{\rho_1\rho_3}\cR_{\rho_4\rho_3}^{\nu\,\rho_5}
\cR_{\rho_1\rho_5}^{\rho_4\rho_2}
\,.
\eeqa
%
%

%%%%%%%%%%%%%%%%%%%%%%%%%%%%%%%%%%%%%%%%%%%%%%%%%%%%%%%%%%%%%%%%%%%%%
%%%%%%%%%%%%%%%%%%%%%%%%%%%%%%%%%%%%%%%%%%%%%%%%%%%%%%%%%%%%%%%%%%%%%

\end{document}